# Incorporation of $H_2$ in vitreous silica, qualitative and quantitative determination from Raman and infrared spectroscopy

B. C. Schmidt[*], F. M. Holtz[1] and J. -M. Bény

Centre de Recherches sur la Synthèse et la Chimie des Minéraux, CNRS, F-45071 Orléans, France – Devenu ISTO – UMR6113

## Abstract

Incorporation mechanisms of $H_2$ in silica glass were studied with Raman and infrared (IR) microspectroscopy. Hydrogenated samples were prepared at temperatures between 800°C and 955°C at 2 kbar total pressure. Hydrogen fugacities ($f_{H2}$) were controlled using the double capsule technique with the iron–wüstite (IW) buffer assemblage generating $f_{H2}$ of 1290–1370 bars corresponding to $H_2$ partial pressures ($P_{H2}$) of 960–975 bars. We found that silica glass hydrogenated under such conditions contains molecular hydrogen ($H_2$) in addition to SiH and SiOH groups. $H_2$ molecules dissolved in the quenched glasses introduce a band at 4136 cm$^{-1}$ in the Raman spectra which in comparison to that of gaseous $H_2$ is wider and is shifted to lower frequency. IR spectra of hydrogenated samples contain a band at 4138 cm$^{-1}$ which we assign to the stretching vibration of $H_2$ molecules located in non-centrosymmetric sites. The Raman and IR spectra indicate that the dissolved $H_2$ molecules interact with the silicate network. We suggest that the $H_2$ band is the envelope of at least three components due to the occupation of at least three different interstitial sites by $H_2$ molecules. Both, Raman and IR spectra of hydrogenated glasses contain bands at ~2255 cm$^{-1}$ which may be due to the vibration of SiH groups. Under the assumption that the reaction Si–O–Si + $H_2$ → Si–H + Si–O–H describes adequately the 'chemical dissolution' of $H_2$ molecules, the SiH concentrations in our samples were determined and the molar extinction coefficient for the SiH absorption band in the infrared ($\varepsilon_{2255}$(SiH)) could then be estimated to be 45 ± 3 l/mol cm. The solubility of molecular $H_2$ in our hydrogenated samples was determined using the IR absorption band at 4138 cm$^{-1}$ and the extinction coefficient given by Shelby [J. Non-Cryst. Solids 179 (1994) 138]. Samples quenched with different cooling rates gave nearly identical Raman and IR spectra, suggesting that the chemical dissolution of hydrogen (SiH and SiOH) can be quenched to room temperature without changing relative concentrations and that no exsolution of hydrogen occurred during the quench.

# 1. Introduction

The incorporation of volatiles such as noble gases, carbon dioxide and water in silica glasses and melts has been widely investigated [2, 3, 4, 5, 6, 7, 8] with respect to physical properties affected by the presence of the volatiles and their related species in the silicate network (see [9] for a recent review of microscopic models for water dissolution). For example, water was shown to decrease melting temperatures [2, 10, 11] and viscosities [12], to increase devitrification rates [13] and to affect optical properties of vitreous silica [14, 15]. Therefore, the dissolution mechanisms of water have been the object of various studies using vibrational spectroscopic techniques such as Raman and infrared (IR) spectroscopy (see [16, 17] for recent reviews).

Molecular hydrogen ($H_2$) was also shown to affect the optical properties of silica glass. Faile and co-workers [18, 19, 20] showed that hydrogen impregnated silica glasses suppress the formation of color centers caused by neutron and gamma irradiation. Dissolution mechanisms of $H_2$ in vitreous silica were studied on samples quenched from silica liquids ($T > 1750°C$) molten in presence of $H_2$-containing atmospheres [1, 21], on silica glasses impregnated with $H_2$ at temperatures below 225°C and hydrogen pressures ranging from 1 to 860 bars [1, 22, 23, 24] or on samples hydrogenated at $P_{H2}=1$ bar and temperatures up to 900°C [25]. These studies showed that hydrogen can dissolve 'physically' as molecular $H_2$ in the interstices of the silicate network, but also 'chemically' due to the dissociation of $H_2$ molecules and the formation of hydroxyl groups (OH) and hydrides (SiH). However, SiH and OH groups were not always observed in hydrogen treated samples and their formation seems to be correlated rather with high temperatures than with high hydrogen pressures prevailing during hydrogenation. Most of the studies on hydrogen dissolution mechanisms used Raman or IR spectroscopy and these two methods can provide complementary information about different H-bearing species dissolved in a glass. To our knowledge, no combined IR and Raman spectroscopic studies for hydrogenated silica can be found in the literature. Therefore, the comparison of previously published Raman and IR spectra suffers from the fact that the spectra were obtained from different silica samples, hydrogenated at different conditions ($P_{H2}$ and $T$). In addition, the spectroscopic studies were often focused on a limited frequency range, not covering the entire range in which the signals of SiH, OH and $H_2$ can be observed (2000–4200 cm$^{-1}$).

The aim of this study is to provide information about the incorporation mechanisms of molecular $H_2$ in silica glass hydrogenated at temperatures 800°C and 955°C and $H_2$ pressures, 960–975 bars. Additionally, experiments with different quench rates have been performed to study the possible effects of different cooling rates on hydrogen solubility in vitreous silica. Such quench effects were indeed observed for the water speciation in hydrous silicate glasses [26], in which the ratio of structurally bonded OH groups and molecular water decreases with decreasing quench rates of the samples. Hydrogenated silica glasses for our study were prepared using the double capsule technique of Eugster [27] and the iron–wüstite–$H_2O$ (IW) oxygen buffer assemblage to control $f_{H2}$ during the experiments. Both, Raman and IR spectroscopy were used for the investigations of our samples providing qualitative and quantitative information about hydrogen solubility in silica glass.

# 2. Experimental procedures

## 2.1. Starting materials

The starting glass for the hydrogenation experiments was a bubble free specimen of Quartzil C (Metaceram – Quartex company, Villejuif, France). Its water content was determined from IR band heights of the fundamental OH-stretching band around 3673 cm$^{-1}$ using the modified law of Bouguer–Lambert–Beer [28]:

$$c = \frac{\text{mw}\, A}{d \rho \varepsilon},$$

where $c$ is the concentration expressed as weight fraction, mw is the molecular weight of the absorbing species, $A$ is the absorbance expressed as peak height, $d$ is the sample thickness (cm), $\rho$ is the sample density (g/l) and $\varepsilon$ is the molar extinction coefficient (l/mol cm). Under the assumption that at very low water content (<0.2 wt%) only O–H vibrations of structurally bonded hydroxyl groups (SiOH) contribute to this absorption band and using the molar extinction coefficient published in [17] ($\varepsilon_{3673}$(OH)=77.5 l/mol cm) the OH content was determined to be 192 ± 10 ppm (in wt% OH).

The metallic iron of the buffer assemblage controlling $f_{H2}$ was a commercial product (Merck LAB) and 'FeO' was synthesized from $Fe_3O_4$ in a $f_{O2}$-controlled furnace at 1 atm and 1200°C (details are given in [29]).

## 2.2. Sample preparation

The starting glasses were cut with a diamond saw into rectangular pieces, typical sizes were 4 × 1.5 × 1.5 mm. To prevent puncturing the capsule under pressure, the sharp edges of the sawn blocks were rounded with a diamond file. After this treatment, the glasses were cleaned ultrasonically in acetone and $H_2O$. After drying for several hours at 110°C in air, the glass blocks were loaded into Pt or Au capsules (10–15 in mm length; 2.5 mm inner diameter; 0.2 mm wall thickness) which were welded shut. The glass-containing capsules were placed together with the oxygen buffer assemblage into Au capsules (30–40 mm length; 5.0 mm inner diameter; 0.4 mm wall thickness) that were also welded at both ends. The solid phases of the buffer assemblages consisted always of 99 wt% reduced (Fe) and 1 wt% oxidized phases (FeO), the amount of $H_2O$ added was always sufficient to oxidize the entire assemblage. In one case the $H_2O$ of the buffer assemblage was replaced by $D_2O$ to obtain further spectroscopic information about the incorporation mechanism of hydrogen into the silica glass. The sealed capsules were tested for leakage by weighing before and after heating at 110°C for 15–30 min.

Hydrogen fugacities imposed by the buffer assemblages were calculated from oxygen fugacities obtained from O'Neill and Pownceby [30] assuming ideal mixing of real gases (Lewis–Randall rule) in the $H_2O$–$H_2$ fluid mixture of the buffer assemblage (Table 1). A more detailed description and discussion of the applied calculations is given in [29]. Buffer masses to be used for a given experimental conditions were calculated on the basis of the $f_{H2}$ imposed by the buffer assemblage and the theoretical mass transfer of $H_2$ through the capsule walls toward the pressure vessel (Eq. (10) in [31]). Hydrogen permeability constants for noble metals and buffer capacities of the oxygen buffer assemblages were taken from [32].

Table 1. : Experimental conditions of hydrogenation experiments

| Run no. | P (kbar) | T (°C) | Duration [a] (min) | $fH_2$ (bar) | $PH_2$ (bar) | Quench | Capsule material |
|---|---|---|---|---|---|---|---|
| Q2-IW [b] | 2 | 800 | 600 | 1370 | 974 | SQ | Pt |
| Q3-IW [b] | 2 | 805 | 600 | 1370 | 974 | RQ | Pt |
| Q5-IW-D2 | 2 | 955 | 90 | 1292 | 962 | RQ | Pt |
| Q6-IW | 2 | 955 | 90 | 1292 | 962 | RQ | Au |

Notes: $f_{H_2}$ and $P_{H_2}$ calculated using the Lewis Randall rule; $D_2 = H_2O$ of the buffer assemblage was replaced by $D_2O$; RQ = rapid quench; SQ = slow quench.
[a] Experimental duration after attainment of the desired P and T.

## 2.3. Apparatus and experimental conditions

The experiments were performed in an internally heated pressure vessel (IHPV), equipped with the fast-quench device of Roux and Lefèvre [33] which was modified for this study. In contrast to the original setup the sample capsules were placed in a sample holder (alumina tube 50 mm in length, 10 mm inner diameter and 0.6 mm wall thickness); the free volume of this tube was filled with silica wool. The sample holder was suspended by a quench wire (Kanthal, 0.2 mm diameter), connected to two electrodes. At the end of the experiment the electrodes were connected to a variac and the voltage was increased which led to electrical fusion of the quench wire and the sample holder dropped into the cold bottom part ($T < 50°C$) of the vessel. The cooling time to room temperature for a quenched sample is estimated to be less than 1 min [33] with high quench rates of 100–50°C/s at the beginning and decreasing quench rates near the end of the cooling. During the experiment, the suspended sample holder was located in the hot spot zone of the furnace, where the thermal gradient could be minimized to less than 5°C by adjusting the power supplies of two windings. Temperature was recorded by three sheathed chromel–alumel thermocouples, calibrated at 1 atm against the melting points of LiCl and NaCl (temperatures are considered to be accurate to ±5°C). Pressure was recorded by a transducer, calibrated against a Heise tube gauge (considered to be accurate to ±20 bars).

Hydrogenation experiments were performed at 2 kbar total pressure and 800°C and 955°C (Table 1). These temperatures were below the glass transition temperatures $T_g$ ($T_g \approx 1205°C$ [34]), allowing only $H_2$ incorporation mechanisms into glasses (not into liquids) to be studied. The experimental conditions were a compromise between the P–T limitations of the experimental apparatus and the buffer lifetimes which decrease rapidly with increasing temperature. Run duration were 1.5 h after reaching the final P and T for the 955°C experiments and 10 h for the 800°C experiments. In most cases quenching was performed using the fast quench device described above, but one experiment was quenched at a smaller rate by turning off the power while the sample remained in the hot part of the vessel. In this case temperature dropped from 800°C to 400°C within 7 min and from 400°C to 100°C within 26 min. After quenching only the external capsules were swollen (due to the high $P_{H_2}$ in the $H_2O$–$H_2$ fluid) but not the sample-containing capsules since no free fluid phase was present during the experiments. Buffer assemblages were checked after the experiments for the residual phases by weight loss ($H_2$, $H_2O$) and X-ray diffraction (Fe, FeO). After the 955°C experiments all phases were present after the quench, demonstrating that the buffer was working until the end. After the long duration experiments at 800°C the buffer contained mainly FeO indicating that the buffer was consumed before the end of the run. However, the external capsules were still swollen which showed that the $P_{H_2}$ at the end of the experiment was only slightly below that of the equilibrium value for the IW buffer assemblage. Doubly polished sections for the spectroscopic investigations were prepared by cutting the quenched samples into plates with a diamond saw and polishing them to sections of 0.5–1.0 mm thickness.

### 2.4. Analytical techniques

Infrared absorption spectra were obtained in the spectral range 600–6000 cm$^{-1}$ using a Fourier transform (FT)–IR spectrometer (Nicolet$^{TM}$ 710) and an infrared microscope (Nicplan$^{TM}$), both purged with dry air. Spectra were obtained using a Globar light source, KBr beamsplitter and a MCT (HgCdTe) detector. The microscope was equipped with a cassegrainian objective (15×), the analyzed spot had a diameter of 100 μm. The samples were positioned on a NaCl disk, the beam was focused into the center of the doubly polished samples. Spectra were acquired in 100 scans and 2 cm$^{-1}$ spectral resolution was consistently chosen.

Raman spectra were obtained using a petrographic microscope (Olympus$^{TM}$) attached to a confocal micro-Raman spectrometer (Dilor$^{TM}$ XY) equipped with a premonochromator and a charge-coupled detector (CCD, Wright$^{TM}$). The samples were excited with the 514 nm monochromatic light of an Argon laser (Coherent Innova$^{TM}$). The laser power was chosen to be 500 mW, the laser beam was focused 30 μm into sample (objective 50×, 8 mm working distance), the confocal aperture was opened to 500 μm. Complete spectra were taken with a resolution between 2.3 and 1.4 cm$^{-1}$ (1200 g/mm holographic grating) in the spectral ranges 235–2300, 1547–3346 and 2764–4331 cm$^{-1}$. Spectra with higher resolution (0.3–0.55 cm$^{-1}$; 1800 g/mm holographic grating) were collected in the spectral ranges of interest. Acquisition times ranged from 50 to 600 s. Identical acquisition parameters were used for spectra that were to be compared. Polished samples were always used to reduce the experimental error. For example, spectra collected from the same sample on different days, but using exactly the same operating parameters, were nearly identical with intensity variations of less than 5%.

## 3. Results

The spectroscopic analysis of the quenched glasses was performed as soon as possible after synthesis and polishing (delay between experiment and analysis was less than 5 days) because of the possible H$_2$ loss from silicate glasses, observed in previous studies. Faile and Roy [20] reported that as much as half of the hydrogen initially present in vitreous silica had diffused out of the glass after 100 days. We observed a similar loss in which 35% of H$_2$ was lost from our samples (polished sections, around 1 × 1.5 × 1.5 mm in size) after 8 weeks.

Typical Raman spectra obtained from vitreous silica before and after hydrogenation are shown in Fig. 1. The higher frequency region (Fig. 1(a)) of the spectrum of the hydrogenated sample (Q6-IW) contains a band at 4136 cm$^{-1}$, which is attributed to the stretching mode of molecular H$_2$ dissolved in the glass [21, 22]. At around 3700 cm$^{-1}$, a band can be detected which can be assigned to O–H stretching vibrations of SiOH groups (e.g. [35]). The asymmetrical band at 2255 cm$^{-1}$ (Fig. 1(b)), present in the hydrogenated glass has been previously detected in the Raman spectra of H$_2$-bearing silica glass [21] and was attributed to Si–H stretching vibrations. The band at 2327 cm$^{-1}$ (located here on the high frequency tail of the SiH band) is attributed to molecular N$_2$ of the air in the laboratory and is present in every Raman spectrum, collected in this frequency range. The low frequency region (Fig. 1(c)) of our samples changes slightly upon incorporation of H$_2$ and its related species. The 'defect bands' at 600 and 490 cm$^{-1}$ (attributed to various structural units such as rings of 3- and 4-

membered SiO$_4$-tetrahedra [36], broken Si–O bonds [37] or overcoordinated Si and O [38]) decrease in amplitude with the incorporation of hydrogen into the glass. The amplitude decrease of the 600 cm$^{-1}$ band is larger than that of the 490 cm$^{-1}$ band.

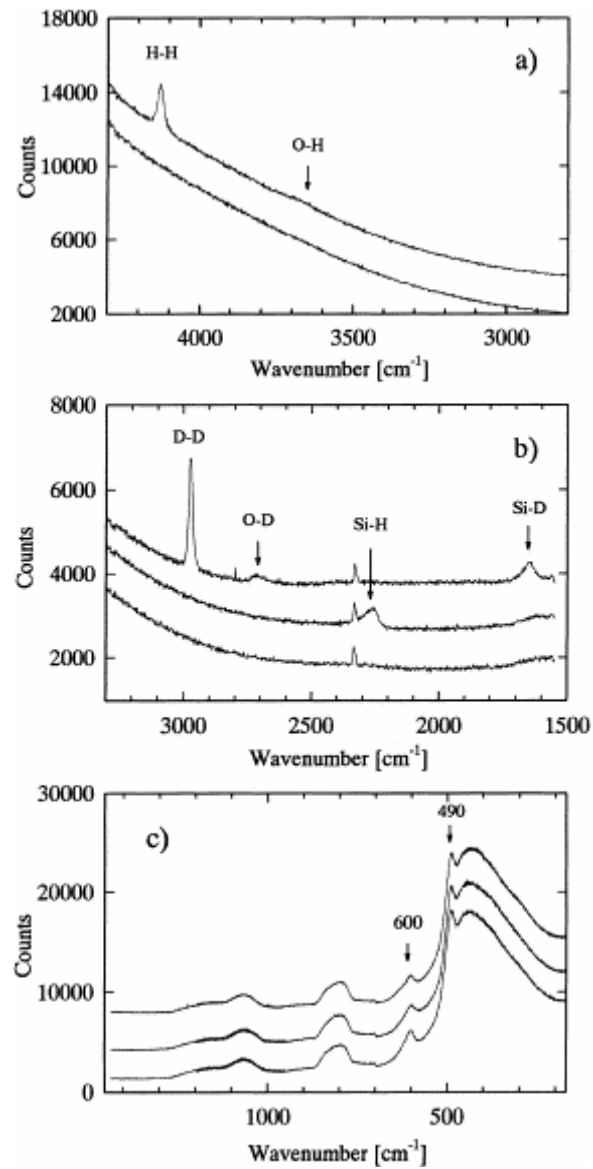

Fig. 1. : Raman spectra of a vitreous silica sample before and after hydrogenation or deuteration. The complete frequency range investigated (4300–170 cm$^{-1}$) is divided in three frequency regions: (a) 4300–2800 cm$^{-1}$, (b) 3300–1500 cm$^{-1}$, (c) 1450–170 cm$^{-1}$. Spectra of the starting sample (Qz(I)) are always plotted below the spectra of the deuterated sample (Q5-IW-D$_2$, on top) and the hydrogenated sample (Q6-IW, middle). Spectral changes due to H$_2$ or D$_2$ incorporation are indicated with arrows and are discussed in the text. The narrow band at 2327 cm$^{-1}$ (b) is assigned to molecular N$_2$ of the laboratory atmosphere and is present in each Raman spectrum collected in this frequency region.

Profiles taken from the centers to the edges of hydrogenated specimen showed no variation in the amplitudes of H-related bands in Raman or IR spectra. Fig. 2 shows the infrared spectra in the frequency range between 2000 and 5500 cm$^{-1}$ of the same silica samples presented in Fig. 1. The spectra (before and after H$_2$ or D$_2$ treatment) are normalized to a sample thickness of 1.0 mm. IR spectroscopy, which is sensitive to the OH stretching vibration (here around 3673 cm$^{-1}$ in Fig. 2(a)), shows that more OH groups are present in the hydrogenated sample than in the starting glass (amplitude increase of the 3673 cm$^{-1}$ band). Although silica glass possesses an infrared absorption band at ~2252 cm$^{-1}$, which is not related to the presence of H-bearing species (see H–D exchange experiments), the appearance of a second band at 2255 cm$^{-1}$ was observed in the hydrogenous sample (Fig. 2(a)). This band was previously attributed to the vibration of SiH groups [19]. The band around 4520 cm$^{-1}$ is assumed to consist of two different contributions: (1) the first SiH overtone (2 × 2255 cm$^{-1}$) and (2) the SiOH combination band (O–H and Si–OH). However, the OH concentration in this glass is less than 750 ppm and the contribution of the SiOH combination band will be very small compared to that of the SiH overtone. The IR spectra of hydrogenated samples show also a band at 4138 cm$^{-1}$ attributed to molecular hydrogen [1, 39]. Because of the centrosymmetry of diatomic, homonuclear molecules, the vibrations of undisturbed H$_2$ molecules are not IR-active and such a band should not be present in the IR spectra. Freund et al. [40] attributed a band at 4130 cm$^{-1}$ in the infrared spectra of crystalline CaO, to the H–H stretching mode of H$_2$ molecules located in non-centrosymmetrical sites. A similar effect may be responsible for the H–H vibration band in the IR absorption spectra of our samples, implying that H$_2$ molecules are bonded to the silicate framework or are distorted by the asymmetric field within the interstices of the silicate network.

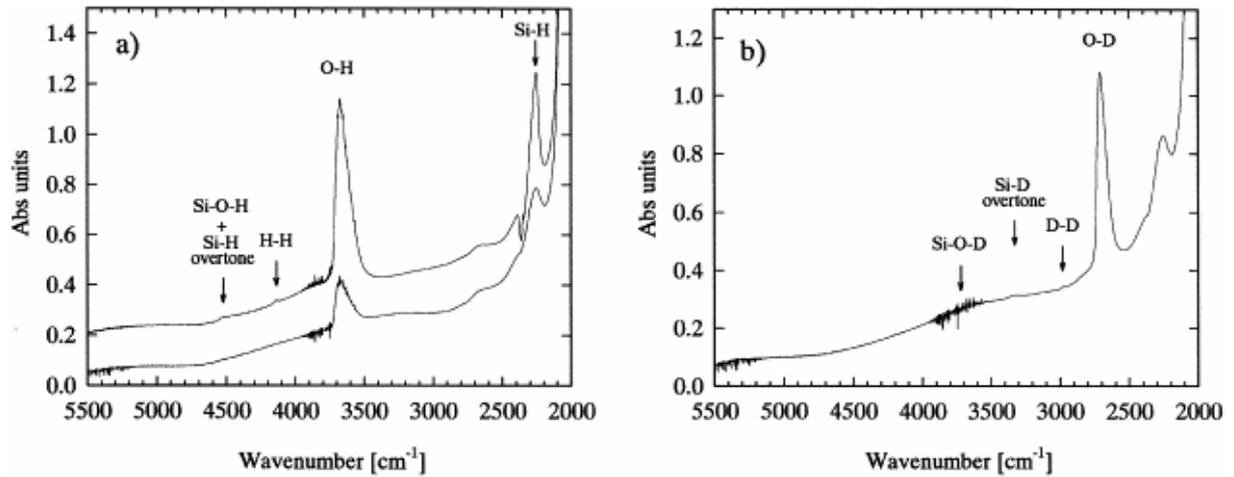

Fig. 2. : (a) IR spectra of a silica sample before (lower spectrum) and after hydrogenation (upper spectrum). For comparison both spectra were normalized to the same sample thickness (1.0 mm). Spectral changes after hydrogenation are indicated by arrows. The negative absorption doublet around 2350 cm$^{-1}$ and the background scatter around 3750 and 5350 cm$^{-1}$ are due to molecular CO$_2$ and H$_2$O, respectively, in the laboratory atmosphere and are not intrinsic features of the samples. (b) IR spectrum of a deuterated sample (Q5-IW-D$_2$). Absorption bands of D-bearing species are indicated with arrows. The SiOD combination band (~3670 cm$^{-1}$) is a smaller feature and is lost in the background scatter due to atmospheric H$_2$O.

The assignments of bands to H-bearing species can be verified by the isotopic replacement of hydrogen by deuterium. The resulting frequency change in the vibrational spectra follows the relation [41]

$$\nu_{XD} = \nu_{XH}\sqrt{\frac{\mu_{XH}}{\mu_{XD}}},$$

under the assumption, that the force constant of the vibration is unaffected by the substitution. In Eq. 2, $\nu$ is the vibration frequency, $\mu$ the reduced mass ($\mu=(m_a m_b)/(m_a + m_b)$, where $m_a$ and $m_b$ are the masses of the two atoms involved in the vibration) and X=Si, O, H, or D.

The Raman and IR spectra of a deuterated sample are given in Fig. 1 and Fig. 2(b), respectively. As predicted, the bands attributed to H–H, O–H and Si–H vibrations are shifted to lower frequencies with the substitution of H by D, confirming the band assignments made above. However, the calculated frequencies for the vibrations of D-containing species are consistently lower (20–50 cm$^{-1}$) than the observed band positions (Table 2). This shift may be due to the expression for the reduced mass in Eq. 2 which is strictly valid only for diatomic molecules. However, the H- and D-bearing species dissolved in our samples are either a part of the silicate network (SiOH, SiOD, SiH and SiD) or have interactions with it (H$_2$ and D$_2$) which may affect the reduced mass of the vibrating groups. Therefore, the expression $\mu=(m_a m_b)/(m_a + m_b)$ applied in our calculations may only be an approximation, leading to the differences between calculated and observed frequencies (Table 2). Although hydrogen was replaced completely by deuterium in all H-bearing species, the IR spectrum of the deuterated sample (Fig. 2(b)) contains an absorption band at 2252 cm$^{-1}$, demonstrating that this band is not due to H-bearing species, but is related to the silicate network of the silica sample (the background scatter in the frequency regions around 3700 and 5350 cm$^{-1}$ is due to atmospheric H$_2$O and is not related to H$_2$O dissolved in the sample). For the deuterium containing glass, the SiD overtone at ∼3350 cm$^{-1}$ can be isolated from the SiOD combination band which is located at ∼3670 cm$^{-1}$. However, in the presented spectrum the latter cannot be resolved from the background scatter in this region (due to atmospheric H$_2$O) since the SiOD concentration is very low.

Table 2. : Calculated and observed frequencies for the vibrations of D-bearing species

| Vibration | $(\mu_{XH}/\mu_{XD})^{1/2}$ | $\nu_{XD}$ calculated (cm$^{-1}$) | $\nu_{XD}$ observed (cm$^{-1}$) |
|---|---|---|---|
| D–D | 0.7075 | 2926 ± 1.4 | 2973 ± 2 |
| O–D | 0.7281 | 2676 ± 1.5 | 2711 ± 2 |
| Si–D | 0.7196 | 1623 ± 1.4 | 1640 ± 2 |

Abbreviations: X = H, D, O, Si; $\mu$ = reduced mass of the vibrating group ($\mu=(m_a m_b)/(m_a + m_b)$).

# 4. Discussion

## 4.1. Comparison of gaseous H$_2$ versus H$_2$ dissolved in silica glass

The high frequency Raman spectrum of gaseous H$_2$ (Fig. 3) consists of four narrow $Q$-branch bands ($Q_1(J)$ with $\Delta\nu=1$; $\Delta J=0$; $J=0,1,2,3$), which are due to vibrational–rotational transitions [42]. In the low frequency region (inset Fig. 3), molecular H$_2$ gives rise to four pure rotational bands ($S_0(J)$ with $\Delta\nu=0$, $\Delta J=2$, $J=0,1,2,3$) and their intensities reflect the populations of the initial rotational levels. H$_2$ molecules exist in two forms with different nuclear spin states, (1) *ortho*-hydrogen with parallel nuclear spins and (2) *para*-hydrogen with anti-parallel spins. In contrast to *para*-hydrogen, *ortho*-hydrogen cannot exist in a state of $J=0$ and its rotational ground state will be $J=1$. Because of the larger abundance of *ortho*-hydrogen (normal population ratio of *ortho*- and *para*-hydrogen: 3:1), the bands which are due to transitions involving rotation energy levels with $J=1$ ($Q_1(1)$, $S_0(1)$) have the highest intensity in the Raman spectra of gaseous H$_2$ (Fig. 3). In contrast, hydrogen molecules dissolved in our samples show in the high frequency region only one band with a larger width and the maximum at 4136 cm$^{-1}$. The rotational fine structure of the (rotational-) vibrational bands observed in the spectra of gaseous H$_2$ cannot be resolved in the spectra of dissolved H$_2$ or has been lost due to the dissolution process. In the lower frequency region (e.g. Fig. 1(c)), the signal of pure rotational transitions ($S_0(J)$ in Fig. 3) disappeared with incorporation of $H_2$ into the glass. The loss of the rotational structure in the $H_2$ spectrum, the frequency shift of the stretching vibration to lower frequency (4136 cm$^{-1}$ in comparison to the $Q_1(1)$ frequency of 4155 cm$^{-1}$ of gaseous $H_2$ at 1 atm [42]), and the bandwidth increase can be assigned to interactions between the dissolved $H_2$ molecules and the surrounding atoms of the silicate network. However, this interaction can be only very weak since $H_2$ molecules diffused out of the hydrogenated samples while stored at room temperature.

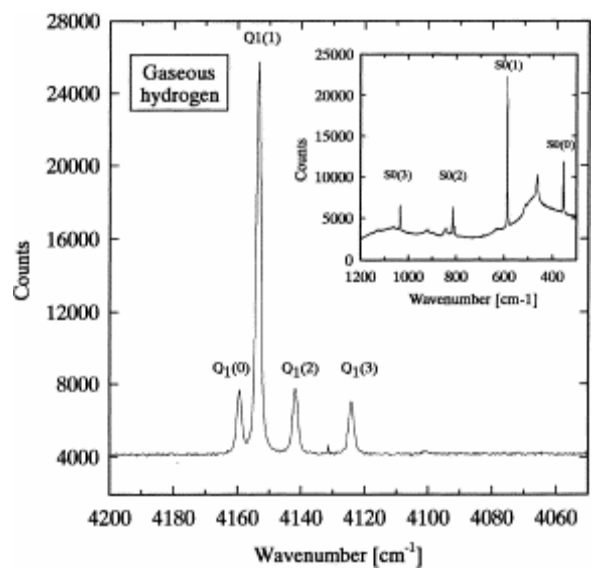

Fig. 3. : Raman spectrum of gaseous $H_2$ trapped in a fluid inclusion with the typical Q-branch bands ($Q_1(J)$) of rotational–vibrational transitions in the frequency range 4160–4120 cm$^{-1}$. The inset shows the pure rotational bands ($S_0(J)$) of gaseous $H_2$, located in the low frequency region of the Raman spectrum.

The presence of the $H_2$ band in the infrared spectra of hydrogenous silica samples indicates a perturbation in the symmetric distribution of the electron density of the $H_2$ molecules under the affects of the surrounding silicate network. The shapes of the $H_2$ stretching bands are nearly similar in the Raman and IR spectra (Fig. 4), suggesting that all dissolved $H_2$ molecules are equally perturbed, i.e. they possess a similar deviation from centrosymmetry. The shape of the $H_2$ band in the spectra of our samples is distinctively asymmetric with a steeper tail at the low frequency side and a small shoulder at around 4110 cm$^{-1}$. We suggest that this asymmetrical shape shows that the $H_2$ band is the envelope of at least three bands implying the occupation of different interstitial sites with different $H_2$–silicate network interactions. Another explanation for the asymmetrical shape of the $H_2$ band could be, that this band is the envelope of the enlarged and frequency decreased $Q$-branch bands of gaseous $H_2$. However, it was impossible to fit this band with four components possessing the properties of the $Q$-branch bands (relative frequencies, peak heights and widths). In addition, the Raman and IR spectroscopic study of hydrogenated silicate glasses ranging in composition from $SiO_2$ to $NaAlSi_3O_8$ [43] showed systematic changes in the shape of the $H_2$ bands along that join, but also between the Raman and IR bands of the intermediate compositions, supporting the suggestion that different interstices are occupied by the dissolved $H_2$ molecules.

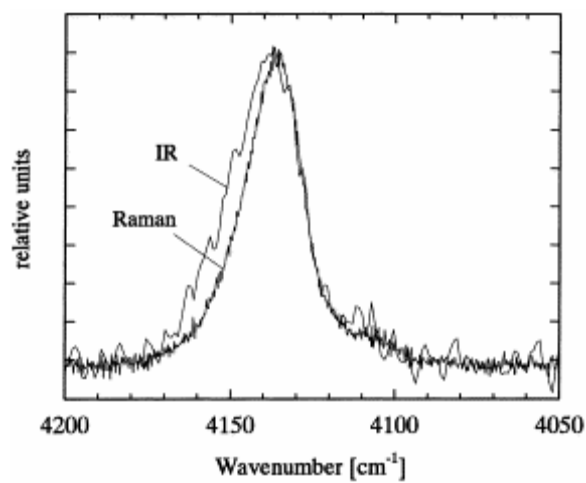

Fig. 4. : Comparison of the $H_2$-bands in the Raman (fine scattered line) and IR spectra (coarse scattered line) of sample Q6-IW. Both spectra are baseline corrected and are normalized to the same peak height.

In the hydrogenated silica samples IR and Raman bands attributed to SiH and OH groups have been observed. These structural units result, most probably, from the 'chemical' dissolution of $H_2$ according to the reaction

$$\text{Si--O--Si} + H_2 \Leftrightarrow \text{Si--O--H} + \text{Si--H},$$

proposed by van der Steen and van den Boom [21]. Shelby [1] has shown that the fractions of hydroxyl and hydride removed from hydrogenated silica during a heat treatment are identical, indicating that SiOH and SiH are most probably formed according reaction (3). The formation of SiH and SiOH groups is probably linked to the presence of certain structural units in the silicate network. Similar to the incorporation of water in fused silica [44], the incorporation of hydrogen or deuterium into silica glass led to the intensity decrease of the 'defect bands' at 600 and 490 cm$^{-1}$ in the Raman spectra (Fig. 1(c)). These bands were assigned to different structural units [36, 37, 38] but the assignments are still under debate. However, we suggest that the Raman spectra show that the first Si–O–Si bonds that break due to $H_2$ incorporation are probably located in the structural units represented by these defect bands. Therefore, the formation of SiH and OH may depend on the type of glass and its thermal history.

## 4.2. Possible quench effects on $H_2$ solubility

An important problem when studying solubility mechanisms of volatiles in silicate systems at elevated pressure and temperature by investigation of the quench products is whether or not the equilibrium species at experimental conditions can be quenched to room temperature. In other words, do the volatiles observed in the glass at room temperature provide information about the volatile species dissolved in the glass or liquid at higher *P* and *T*?

To obtain information about possible quench effects on the H-speciation of hydrogenous glasses, experiments with different quench rates have been performed (Q3-IW, rapid quench: 800 → 50°C in <1 min; Q2-IW, slow quench: 800 → 100°C in 33 min). Although the buffer might have been consumed in these two experiments, information on the effects of different cooling rates can be obtained. Both double capsules were prepared identically and contained the same amounts of the buffer assemblage, assuming that the '$P_{H2}$-history' of both samples was the same. The IR and Raman spectra of both samples (Fig. 5 and Fig. 6) were identical with respect to all observed H-bearing species ($H_2$, OH and SiH), suggesting that neither, physical dissolution ($H_2$ molecules) nor chemical dissolution of $H_2$ (SiH and OH groups) is affected by quench rates or that the equilibrium species in the silicate melt change so rapidly, that it cannot be detected in our experiments. The interactions of dissolved $H_2$ molecules with the surrounding network seem to be indeed temperature dependent. The diffusive mobility of $H_2$ in silicate liquids melts and glasses at high temperatures [45] suggests that these interactions are much weaker at these temperatures than those observed at room temperature. This difference is also indicated by the in situ Raman spectra of hydrogenated silica and aluminosilicate glasses [23, 43] where a shift of the $H_2$ band position towards higher frequency and a change of the bandwidth was observed with increasing temperature. The identical line shapes of the $H_2$ bands in the Raman spectra of glasses quenched with different quench rates is therefore the basis for suggesting that the interactions of $H_2$ molecules with the silicate network cannot be quenched to room temperature.

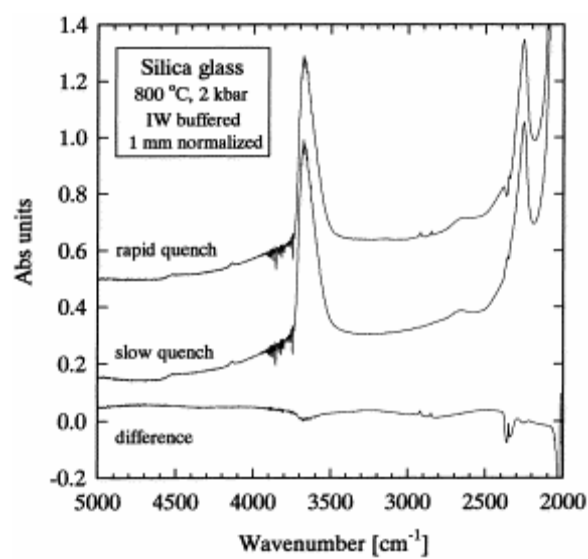

Fig. 5. : IR spectra of samples hydrogenated at 2 kbar, 800°C, IW buffer and quenched with different cooling rates. The two upper spectra show the IR spectra of Q3-IW (rapid quench) and Q2-IW (slow quench), normalized to a sample thickness of 1.0 mm. The lower spectrum shows the difference spectrum (Q3-IW–Q2-IW). No quench effects on the IR bands of H-bearing species in hydrogenated samples ($H_2$, OH, SiH) were observed. The weak negative peaks around 3675 and 2255 cm$^{-1}$ in the difference spectra are assumed to result from non-perfect thickness normalization. The two narrow bands around 2900 cm$^{-1}$ (Q3-IW and difference spectra) result from residual organic material used in the preparation of the sample for the spectroscopic investigations.

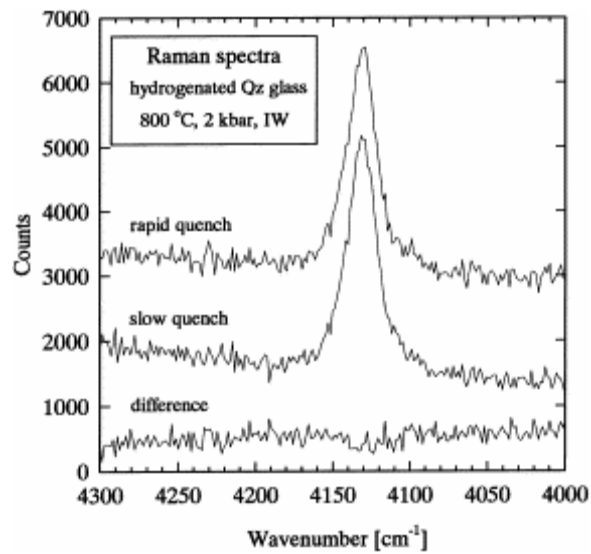

Fig. 6. : Raman spectra of samples hydrogenated at 2 kbar, 800°C, IW buffer and quenched with different cooling rates. The two upper spectra show the IR spectra of Q3-IW (rapid quench) and Q2-IW (slow quench). The lower spectrum shows the difference spectrum (Q3-IW–Q2-IW). No quench effect on the $H_2$ peaks assigned to the H–H stretching vibration in the Raman spectra of hydrogenated samples was observed.

Another possible problem which can occur during quenching of volatile-bearing systems from high $P$ and $T$ is the exsolution of volatile species. However, profiles collected at room temperature from the edges to the centers of hydrogenated silica glass blocks showed no variation of the amounts of H-bearing species. In addition, scanning electron microscope analysis did not detect submicroscopic bubbles or cracks which could be formed during cooling, nor was a gas phase observed in the capsules after quenching indicating that none or very little $H_2$ exsolution occurred during the quench. Furthermore, the spectroscopic investigations showed only the signals of $H_2$ without rotational freedom but no signals of $H_2$ molecules that could be present in liquid-like pools.

Possible problems for quenching SiOH and SiH may also occur. Variations of the proportions of molecular water and hydroxyl as a function of the quench rate have indeed been observed in hydrous aluminosilicate glasses [26]. Analogous to the incorporation of water, the formation of SiOH and SiH groups due to $H_2$ incorporation involves bond breaking and reorganization of the silicate network. This process should be more sluggish than changes in the interaction of $H_2$ molecules with the silicate network and it should be possible to observe it with quench rate dependent experiments. Since we do not observe variations of the relative amounts of $H_2$, SiOH and SiH groups as a function of quench rates, we conclude that the species resulting from chemical dissolution of $H_2$ could be quenched to room temperature. However, it is worth noting that our hydrogenation experiments were conducted on glasses at temperature $<T_g$ and that experiments performed at liquidus conditions may have different results.

### 4.3. Quantitative determination of SiH and $H_2$

According to reaction (3), the formation of SiH groups is coupled with the formation of equal amounts of OH groups. If the concentrations of OH groups before and after hydrogenation are known (e.g. by determination with IR spectroscopy), the concentration of SiH groups can be calculated and the extinction coefficient ($\varepsilon_{2255}$(SiH)) for the IR SiH-band at 2255 cm$^{-1}$ can be determined. However, other reactions between silica and hydrogen can also produce hydroxyl and hydride. For example, reactions of $H_2$ with defects in the silicate network like non-bridging oxygen atoms or oxygen vacancies can result in formation of OH or SiH. Also water trapped in the sample capsule (in form of surface adsorbed water on the samples, atmospheric water or $H_2O$ formed from oxygen containing gases during the reducing conditions of our experiments) can dissolve as hydroxyl groups in the samples at our experimental conditions. To use reaction (3) for the determination of $\varepsilon_{2255}$(SiH) it must be shown that this reaction was the dominant process for the formation of SiH and OH groups in our hydrogenated samples. This demonstration can be done by showing the proportionality between SiH and OH and requires samples with different amounts of hydroxyl and hydride. Such samples can be obtained from synthesis at different $H_2$ pressures or by removing some of the hydroxyl and hydride from a hydrogenated sample by a heat treatment. We heated a doubly polished section of sample Q6-IW for 3 h at 900°C in air and measured IR spectra in intervals of 5–60 min in order to determine the SiH and OH removal from the sample. Absorbances of the bands at ∼3673 cm$^{-1}$ (OH) and ∼2255 cm$^{-1}$ (SiH) were measured from difference spectra, obtained by subtraction of the starting glass spectrum from that of a hydrogenated sample (both spectra normalized to the same sample thickness). Therefore, these spectra are due to either the changes due to hydrogenation (unheated samples) or the changes due to hydrogenation and following heat treatment. The concentration of the 'additional' OH groups ($\Delta c_{OH} = c_{OH}$(hydrogenated sample) $- c_{OH}$(starting glass)) was calculated from heights of the 3673 cm$^{-1}$ IR absorption band using the extinction coefficient $\varepsilon_{3673}$(OH)=77.5 l/mol cm [17] and the density of 2,200 g/cm$^3$ for vitreous silica. Fig. 7 shows the absorbance (band heights) of the SiH band in the difference spectra of our samples (hydrogenated only and hydrogenated and heat treated) as function of $\Delta c_{OH}$. The data define a linear relation, demonstrating the proportionality between SiH and OH groups. However, the linear regression of the data does not pass through the origin and intersects the *x*-axis at 23 ppm OH suggesting that this amount of hydroxyl is formed by another process than that described by reaction (3) and has to be taken into account for the calculation of SiH concentration. On the basis of this consideration we calculated the SiH concentrations for our hydrogenated samples (not heat treated) to be 778 to 911 ± 25 ppm (Table 3). Taking all samples into account, the extinction coefficient $\varepsilon_{2255}$(SiH) was determined to be 45 ± 3 l/mol cm, corresponding to 0.58 ± 0.04 times that of hydroxyl in vitreous silica. This number is close to that of Morimoto et al. [46] (0.57 ± 0.05 times that of hydroxyl) but lies above the extinction coefficient determined by Shelby [47] (0.45 ± 0.06 times that of hydroxyl). Although these values for $\varepsilon_{2255}$(SiH) are in the same order of magnitude, the variations are up to ∼25% and can be caused by several factors. The extinction coefficient $\varepsilon_{2255}$(SiH) determined previously [46, 47] and in this study were obtained assuming that reaction (3) describes adequately the formation of hydroxyl and hydride. However, different silica glass types were used in these studies (Vitreosil Opaque [46], Suprasil, Suprasil W, Amersil T-08 [47] and Quartzil C (this study)) which may have contained different amounts of structural defects (such as oxygen vacancies and non-bridging oxygen atoms) changing the relative amounts of SiH and OH groups, which then lead to some of the differences for the determined $\varepsilon_{2255}$(SiH). In addition, using different spectrometers and different data processing procedures (e.g. baseline corrections) can cause variations for IR extinction coefficients [48]. It is worth noting that the accuracy for the absolute magnitude of $\varepsilon_{2255}$(SiH) and thus the calculated concentration of SiH groups depends on the accuracy of $\varepsilon_{3673}$(OH) used for the determination of the water content.

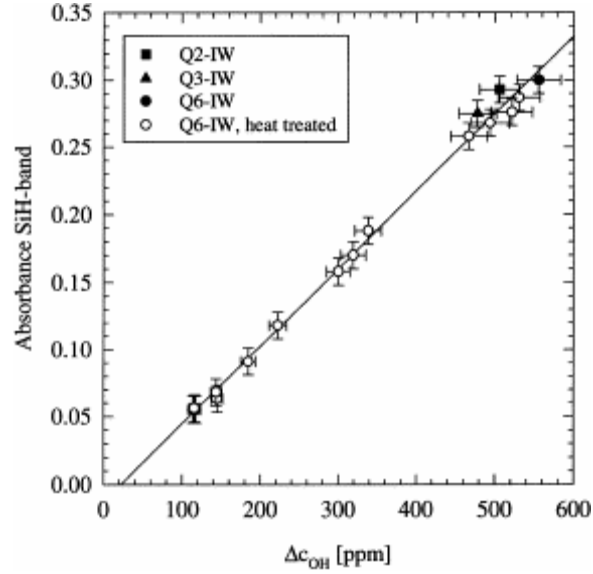

Fig. 7. : Absorbance (band heights) of the SiH band in the difference spectra of our samples (hydrogenated only and hydrogenated and heat treated) as a function of the OH concentration calculated from the OH band at 3673 cm$^{-1}$ in these spectra ($\Delta c_{OH}=c_{OH}$(hydrogenated sample) − $c_{OH}$(starting glass)).

Table 3. : SiH concentration in hydrogenated samples and determination of the extinction coefficient for the IR-band at 2255 cm$^{-1}$, attributed to the vibration of SiH

| Glass | $c_{OH}$ [ppm] | $\Delta c_{OH}$ [ppm] | $c_{SiH}$ [c] [ppm] | Abs$_{2255}$ (SiH) | $\varepsilon_{2255}$ (SiH) [l/mol cm] |
|---|---|---|---|---|---|
| QzI [a] | 192 ± 10 | – | – | – | – |
| Q2-IW [b] | 698 ± 35 | 506 ± 25 | 826 ± 43 | 0.279 ± 0.01 | 46.9 ± 3 |
| Q3-IW [b] | 670 ± 34 | 478 ± 24 | 778 ± 41 | 0.271 ± 0.01 | 46.7 ± 3 |
| Q6-IW | 748 ± 38 | 556 ± 28 | 911 ± 48 | 0.300 ± 0.01 | 43.5 ± 3 |

All concentrations in weight; hydroxyl concentrations were calculated from the peak heights of the 3673 cm$^{-1}$ IR bands, $\rho_{silica}$ = 2200 g/l and $\varepsilon$(OH) = 77.5 l/mol cm [17]; $\Delta c_{OH} = c_{OH}$ (hydrogenated glass) − $c_{OH}$(QzI).
[a] Starting glass.
[b] IW buffer did not work until the end of the experiment.
[c] The formation of ~23 ppm OH due to another process than that described by reaction (3) was taken into account.

The larger amount of additional water and SiH in experiment Q6-IW (2 kbar, 955°C, 2 h, IW) than in experiments Q2-IW and Q3-IW (2 kbar, 800°C, 10 h, IW, see Table 3) is probably not an effect of temperature, but results more probably from problems with the IW buffer assemblages of Q2-IW and Q3-IW, in which the buffer was consumed before the end of the experiments. Therefore $f_{H2}$ at the final stage of the experiments was less than the equilibrium $f_{H2}$ of the IW buffer assemblage.

The amounts of molecular $H_2$ dissolved in our samples were estimated from the heights of the 4138 cm$^{-1}$ bands in the IR spectra and the extinction coefficient $\varepsilon_{4140}$(H–H)=0.26 l/mol cm given in Shelby [1] to range between 349 and 451 ppm $H_2$ corresponding to $2.29 \times 10^{20}$–$2.96 \times 10^{20}$ molecules/cm$^3$ glass. These are less than those of Shelby [1] and Hartwig [22] who reported $H_2$ concentrations of $3.5 \times 10^{20}$ and $6 \times 10^{20}$ molecules/cm$^3$ glass for samples hydrogenated at a $P_{H2}$ of ~860 bars and temperatures of 225°C and 90°C, respectively. Although these numbers should not be compared directly since different starting materials, experimental conditions and analytical techniques were used, the data show unambiguously the negative correlation of $H_2$ solubility with increasing temperature. Such trends were already observed for the solubilities $H_2$, $D_2$ and noble gases in vitreous silica at low pressures (<1 bar) [5, 49] and for Ar at 1200 bars [8].

## 5. Conclusions

For silica glass hydrogenated at 2 kbar total pressure, 800°C and 955°C and $H_2$ pressures of 960–975 bars, two main dissolution mechanisms of $H_2$ have been observed: a 'physical' dissolution of molecular $H_2$ and a 'chemical' dissolution in which $H_2$ dissociates to form SiH and SiOH groups. From the position and the line width of the $H_2$ band in the Raman spectra we conclude that the $H_2$ molecules interact with the silicate network. This interaction is confirmed by the IR spectra showing a band which could be assigned to the vibration of molecular $H_2$. We conclude that in both, Raman and IR spectra the line shape of the $H_2$ band is asymmetrical because it is the envelope of several contributions (at least 3) due to $H_2$ molecules located in different structural sites in the silicate network. The shape of the $H_2$ bands is similar in the IR and Raman spectra, implying that all sites have the same deviation from centrosymmetry. From the results of our quench rate dependent experiments we conclude that the chemical dissolution of hydrogen (SiH and SiOH) can be quenched to room temperature without changing relative concentrations or exsolution of hydrogen. The determined extinction coefficient for the SiH band in the IR spectra of Quartzil C is in reasonable good agreement with those of previous studies, but more data are required to further constrain this coefficient, necessary for reliable determinations of SiH concentrations in vitreous silica.

The spectroscopic results obtained for vitreous silica glass hydrogenated at elevated pressures and temperatures (800°C and 955°C, 2 kbar, IW) are in a good agreement with those of previous studies (performed at either elevated pressures or elevated temperatures). Although the experiments in this study were performed on glasses and not on silica liquids, the results are qualitatively similar to those obtained from glasses which were molten at 1 bar under a pure $H_2$ atmosphere [21].

## Acknowledgements

This research constituted a part of B.C.S.'s PhD thesis, supported by a grant of the French Ministry for Research and Education. The manuscript benefited from the critical comments of two anonymous reviewers.